# Cognitive ability experiment with photosensitive organic molecular thin films

**Régis Barille,[1] Sohrab Ahmadi-Kandjani,[1,2] Ewelina Ortyl,[3] Stanislaw Kucharski[3] and Jean-Michel Nunzi[1]**

[1]*Laboratoire Propriétés Optiques des Matériaux et Applications, UMR CNRS 6136 - Université d'Angers, 2 Boulevard Lavoisier, 49045 Angers, France.*

[2]*Research Institute for Applied Physics and Astronomy, University of Tabriz, 51664 Tabriz, Iran.*

[3]*Institute of Organic and Polymer Technology, Wroclaw Technical University, 50-370, Wroclaw, Poland.*

**ABSTRACT:**

We present an optical experiment which permits to evaluate the information exchange necessary to self-induce cooperatively a well-organized pattern in a randomly activated molecular assembly. A low-power coherent beam carrying polarization and wavelength information is used to organize a surface relief grating on a photochromic polymer thin film which is photo-activated by a powerful incoherent beam. We demonstrate experimentally that less than 1% of the molecules possessing information cooperatively transmit it to the entire photo-activated polymer film.

Information exchange is mandatory to induce self-organization [1-3]. Optical recording and photofabrication techniques use coherent beams and exploit changes in material properties. In this respect, surface relief gratings (SRG) in photosensitive organic molecular thin films have been investigated for their potential applications to optoelectronic devices [4]. In the conventional method to inscribe SRGs in photoactive molecular films, a periodic intensity distribution is produced on the film surface by the interference between two optical beams. The intensity pattern induces isomery, reorientation and molecular migration of the photochromic molecules inside the polymer film [5,6], resulting in a mass transport effect. Spatial patterns may also arise spontaneously as a consequence of nonlinear wave mixing processes in the transverse cross-section of coherent optical beams [7]. As a universal feature, these effects require a coherent excitation which simultaneously carries the information and excites the degrees of freedom of the system: excitation, rotation and translation in the case of photochromic molecules.

We have shown previously that under the assistance of a large-power incoherent beam, the information carried by a low-power coherent beam permits to create a well organized SRG [8]. We seek in this paper for some quantitative insight into the details of the cognitive process: that is the minimum quantity of information exchange leading to self-organization. In this respect, the spatial coherence of the large-power assistance beam is varied from low to high coherence and the diffraction efficiency of the self-induced SRG is monitored. Information in our experiment is given as diffraction characteristics by the coherent signal beam. We show that without coherence, no well defined diffraction pattern is induced. With some degree of coherence, a self induced diffraction pattern is formed more or less rapidly. We thus get a measurement of the information exchange duration in terms of coherence. We also give quantitative insight into the cooperativity of the process. Practically, information given by the signal beam is related to the orientation and pitch of the grating. Our experiments show that photoactive molecular films come into a unique cooperative organization although most of the individual molecules have a random light-assisted movement.

Our samples are polymer films made from a highly photoactive azobenzene derivative containing heterocyclic sulfonamide moieties [9]. The polymer (insert in figure 3) was chosen in virtue of its rather low glass transition temperature Tg = 71.1 °C, leading to fast light induced ripple growth. 1 μm-thick films were deposited on glass substrates by spin-coating of the polymer with a concentration of 50 mg/ml in a THF solution. The $\lambda = 476.5$ nm laser line of a continuous argon ion laser was used to excite the azo-polymer close to its absorption maximum which peaks at 438 nm. Absorbance at the working wavelength is 1.6.

A sketch of our experimental set-up is given in figure 1. The laser power is 500 mW. The beam width is increased to 5.8 mm by an afocal system made of two lenses. The beam is then divided into the two arms of a Mach-Zehnder interferometer. In the first arm, the beam is reduced to a width of 2.1 mm by another afocal. This first beam acts as a signal beam. The second one acts as a totally incoherent or partially coherent pump beam. Incoherence in the second arm is provided by focusing the laser beam through a diffuser (a frosted glass produced by acid etching of a clear sheet glass). By adjusting the position of the diffusing phase screen between the two lenses, we are able to tune the transverse coherence length $\sigma_\mu$ of the pump beam. An additional hot-air gun is used as a turbulent layer generator [10] to generate so-called pseudo-thermal light [11]. It is inserted along the beam path in order to obtain total incoherence and depolarization. The diffuser plus turbulence generator introduce a random phase which varies much faster than the time constant for grating formation in the sample. The two beams illuminate the polymer sample: signal and pump beams overlap spatially at the exit of the Mach-Zehnder, with the thinner signal beam fixed at the center of the pump (Fig.1). Signal beam power is 0.07 mW; its polarization is vertical; diameter is 2.1 mm. Pump beam intensity is 87 mW; its diameter is 5.8 mm.

In a first experiment, we verified that the surface grating stored can not be the result of one beam alone: the signal beam is sent alone onto the sample without the incoherent pump. The signal beam intensity is chosen below the intensity threshold for spontaneous SRG induction in the film. We get a visual proof of spontaneous SRG induction when first order diffraction appears in reflection and can be measured with an appropriate combination of lenses and photodiode. First order diffraction appears with a power threshold of 0.7 mW for the coherent signal beam. No self-structuring of the polymer surface was induced below this threshold after one hour exposure. Following the same verification, when the large 87 mW partially coherent pump beam was sent alone onto the film with only the diffuser placed behind the first lens and without the hot-air gun, no SRG was induced after an even longer exposure time.

When the two overlapping beams are sent together, a SRG is induced after ½ hour exposure without hot air gun, although the maximum incoherence is induced by the diffuser. The diffraction intensity growth shown in figure 2 was measured for different pump beam coherence lengths $\sigma_\mu$ . Coherence length is measured as the speckle size $\sqrt{S/N}$ [12] in which $S$ is the illuminated surface and $N$ the number of speckle spots. Spatial coherence length varies practically from 0 to 1500·λ. Almost complete incoherence and depolarization ($\sigma_\mu / \lambda \approx 0$) is obtained using the hot air gun. In this condition, grating growth rate is the slowest.

For all gratings stored for different pump coherent lengths, the height and pitch of the gratings were retrieved using a contact-mode atomic force microscope (AFM). In all cases and in every region of the illuminated polymer area, the peak to peak height of the grating is $100 \pm 5$ nm; its mean pitch is $\Lambda = 888 \pm 30$ nm; the grating wave-vector is parallel to the polarisation of the signal beam. The pitch $\Lambda$ is close to the theoretical value given by first order diffraction theory in the backward direction with angle $\theta = 32.6°$ from the normal: $\Lambda = 2\lambda / 2\sin\theta$. The pitch is determined by the conditions of maximum diffraction efficiency out of the self-induced grating, when the second order of diffraction occurs along the film plane. This was initially interpreted as "stimulated Wood's anomalies" by A.E. Siegman [13] for similar problems. Importantly, the grating area expands and covers all the illuminated area. The reason is that molecules which move permanently under light reduce their motion only if they find a shadowed region [14]. They thus organize collectively, finding the optimal way to scatter light out of the polymer film. Information about optimal scattering directions: namely the SRG parameters; is given by the coherent beam characteristics: incidence and polarization direction [8, 15]. The molecules which respond to the coherent beam create a diffraction pattern and then transmit the pattern step by step to all their neighbours. Information is exchanged locally via the relief: pits sending more light to the neighboring and dips sending less light, thus replicating the pattern.

The results in figure 2 show that partial coherence of the pump beam accelerates the transfer of information from the signal to all the illuminated area in which self-organization takes place. It is illustrated by plotting the time threshold for self-diffraction of the pump beam as a function of pump coherence length (insert in figure 2). This one yields duration of the information exchange as a function of coherence. Time threshold is taken as the point at which a straight-line fit to the linear portion of the diffraction intensity curve intercepts the X-axis (fig. 2). The result is fitted with the test function $I_t = A / \left(\sigma_\mu / \lambda\right)^B$ in which $A$ and $B$ are constants; the limit $\sigma_\mu \to \infty$ represents a spatially coherent source and $\sigma_\mu \to 0$ an incoherent one.

The learning phenomenon in which the molecules under the incoherent light illumination self-organize according to the coherent beam characteristics is exemplified by another experiment made with a depolarized pump beam. The coherent signal beam is stopped after about one hour of SRG self-organization and only the incoherent beam is left to irradiate the polymer. Figure 3 shows that the diffraction intensity continues to increase, proving that the SRG continues to self-organize. The SRG expands laterally, covering finally all the illuminated surface. It expands step by step due to the short range interference of the diffracted waves with the incident beam. After two hours the hot-air gun is stopped, increasing spatial coherence of the pump. We see that the diffracted intensity increases about 15% in a 2 minutes time lag, without modification of the evolution law. This shows that learning is accelerated by coherence (information exchange) without modification of the information content: namely the SRG characteristics.

In another two-beam experiment, we measured the time evolution of the diffraction intensity when polarization of one of the beams is rotated. The hot air gun is not used and pump polarization is maintained; the pump is then partially coherent. In figure 4, the signal beam polarization is rotated after one hour of interaction (circles). Diffraction which follows the signal beam polarization [16] then changes direction. When the pump polarization is rotated, also after one hour of interaction while keeping the signal beam polarization unchanged (stars in fig.4), diffraction continues to increase after about 10 minutes of time lag. This shows that leadership is acquired by the low power coherent signal beam, whatever its direction of polarization. The pump permits collective SRG growth but does not participate to the decision on the grating characteristics. The time lag after pump polarization rotation corresponds to the time needed for leadership to take place. Practically it corresponds to the time taken by the molecules moving under the partially coherent pump action outside of the signal region to receive the information on the grating characteristics from the signal. This lag may be considered as the minimum learning time in our cognitive process. This last experiment gives further information on the leadership process within azo-polymers: a cooperative decision on the grating characteristics is driven by few informed molecules differing in behaviour from the majority which undergoes almost random molecular migration under the incoherent light exposure.

In order to find a quantitative vision of the cooperativity of the process, we assume that molecules undergo molecular migration when they are brought into the *cis*-state [14]. The coherent signal intensity represents about 0.08% of the incoherent pump power and 13% of the total surface. Let us consider a simplified *trans - cis* isomery model [17]. When the polymer is illuminated by a depolarized background light superimposed to a linearly polarized signal beam with respective intensities $I_b$ and $I_S$, the *cis*-population in the steady state is given by:

$$N_c = N_{to} \frac{q_T \sigma_T (I_S \cos^2 \theta + I_b)}{q_T \sigma_T (I_S \cos^2 \theta + I_b) + q_C \sigma_C (I_S + I_b) + K} \tag{1}$$

where $N_c$ is the concentration of molecules in the *cis*-state under illumination and $N_{to}$ the total number of molecules in the dark; $\theta$ is the average angle between the molecular orientation of the *trans* molecules and the signal beam polarization; $q_T$ and $q_C$ are the quantum yields for *trans*-to-*cis* and *cis*-to-*trans* isomery respectively; $\sigma_T$ and $\sigma_C$ are the absorption cross sections of the *trans* and *cis* forms respectively; $K$ is the thermal relaxation rate from *cis*-to-*trans*. Equation (1) assumes implicitly that molecules in the *cis*-state do not have polarized absorption. We have $N_{to} = 7.18 \cdot 10^{20}$ cm$^{-3}$. We use the following set of parameters: $q_T = 0.1$; $q_C = 0.4$ [18] and $\sigma_T = 0.75 \cdot 10^{-16}$ cm$^{-2}$; $\sigma_C = 0.11 \cdot 10^{-16}$ cm$^{-2}$; $K = 0.094$ s$^{-1}$ [9]. We assume $\theta = 0$ for simplicity. The photon flux is $7.8 \cdot 10^{15}$ cm$^{-2}$s$^{-1}$. We get $N_c = 2.25 \cdot 10^{20}$ cm$^{-3}$ molecules for the steady-state *cis*-concentration under illumination. Derivation of equation (1) with respect to *Is* gives the ratio *r* of molecules which move under the signal influence. We call them the informed ones:

$$r = \left(I_S / N_c\right)\left(\frac{\partial N_c}{\partial I}\right) \approx 0.003 \tag{2}$$

This result illustrates the cooperativity of the process: 0.3 % of the molecules possessing decisive information drive gradually the other molecules into a collective decision. Although the detailed process is different, this high level of cooperativity can be confronted to other experiments borrowed from solid-state physics. For example, a photoinduced phase transitions from the neutral to ionic (ferroelectric) form in an organic charge-transfer crystal was triggered by a tiny femtosecond laser pulse [19]. The phenomenon is triggered by less than one excited molecule for about 100 molecules in the crystal. In another example, cooperative chiral order was induced in a copolymer containing a minority of chiral units (so-called sergeants) among achiral units (so-called soldiers) [20]. The same optical activity as using a pure chiral homopolymer is induced with only 15% of chiral units and substantial activity appears with as low as 0.5% of chiral units. In our experiment, we chose a surface ratio of 7.6 between pump and signal beams for demonstration purposes. Changing this ratio changes the speed of growth but not the final cognitive effect. We expect a threshold when the signal beam size becomes commensurate with the speckle size.

In conclusion we have shown a mechanism of cooperative decision through the process of SRG self-organization. In this work, information about the SRG characteristics (pitch and orientation) is given by a tiny localized coherent signal beam and molecular migration is powered by a strong incoherent or partially coherent pump beam. The rate of information exchange is directly related to the coherence of the pump beam. Cooperative organization outside of the coherent beam area between the informed molecules and the other photoactivated ones is the result of molecular migration which optimizes light scattering away from the polymer film. Learning is achieved step by step with less than 1% of molecules initially informed. Our system should find applications as a paradigm for studies on cognitive ability: various grating parameters can indeed be accessed by tuning the coherent beam characteristics [15]; this is the initial information from which the system finds optimal light scattering efficiency conditions. A practical interest of surface structuring using incoherent light could be that a powerful incoherent light is easier and cheaper to obtain than coherent light.

**FIGURE CAPTIONS**

**Fig. 1**: Scheme of the experimental setup.

**Fig. 2**: First order self-diffraction intensity as a function of time for different values of the beam coherence $\sigma_\mu / \lambda$ given as index to the curves. Coherence of the pump is modified by moving the diffuser between the two lenses. $\sigma_\mu / \lambda \to 0$ corresponds to complete incoherence. Insert shows the time threshold for first order self-diffraction as a function of the transverse coherence length. The curve is fitted with $A / \left( \sigma_\mu / \lambda \right)^B$.

**Fig. 3**: Intensity of first order self-diffraction as a function of time under incoherent pumping. When the signal beam is turned off the SRG continues to grow under the incoherent light excitation. When the hot air is suppressed, the pump beam becomes partially coherent and the SRG grows faster. The dashed line is a parabolic guide to visualize evolution with time. Insert shows the chemical formula of the polymer.

**Fig. 4**: First order self-diffraction intensity as a function of time for the case of a partially coherent pump without hot air gun. When the signal beam polarization is rotated, the SRG changes direction and diffraction which is emitted in a perpendicular direction is no more recorded by the photodiode (circles). When the pump beam polarization is rotated, the surface induced grating continues to grow (stars).

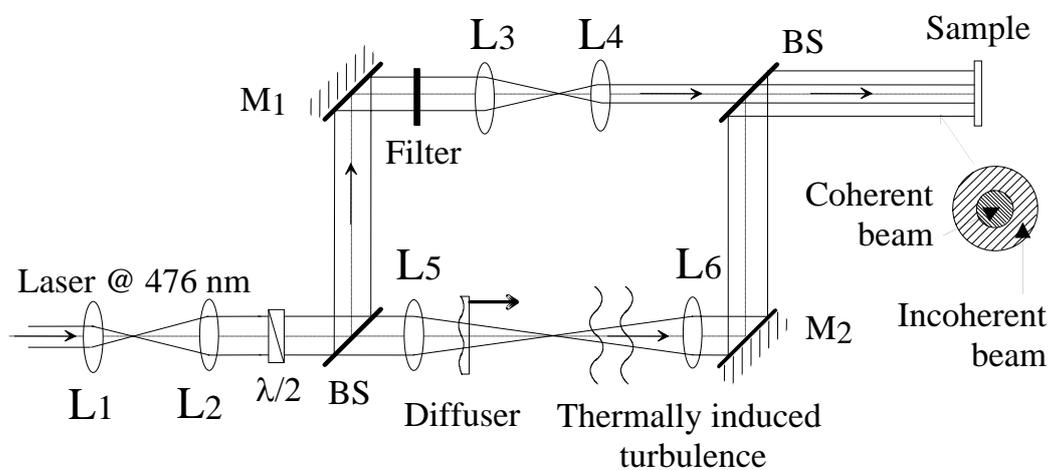

**Figure 1**

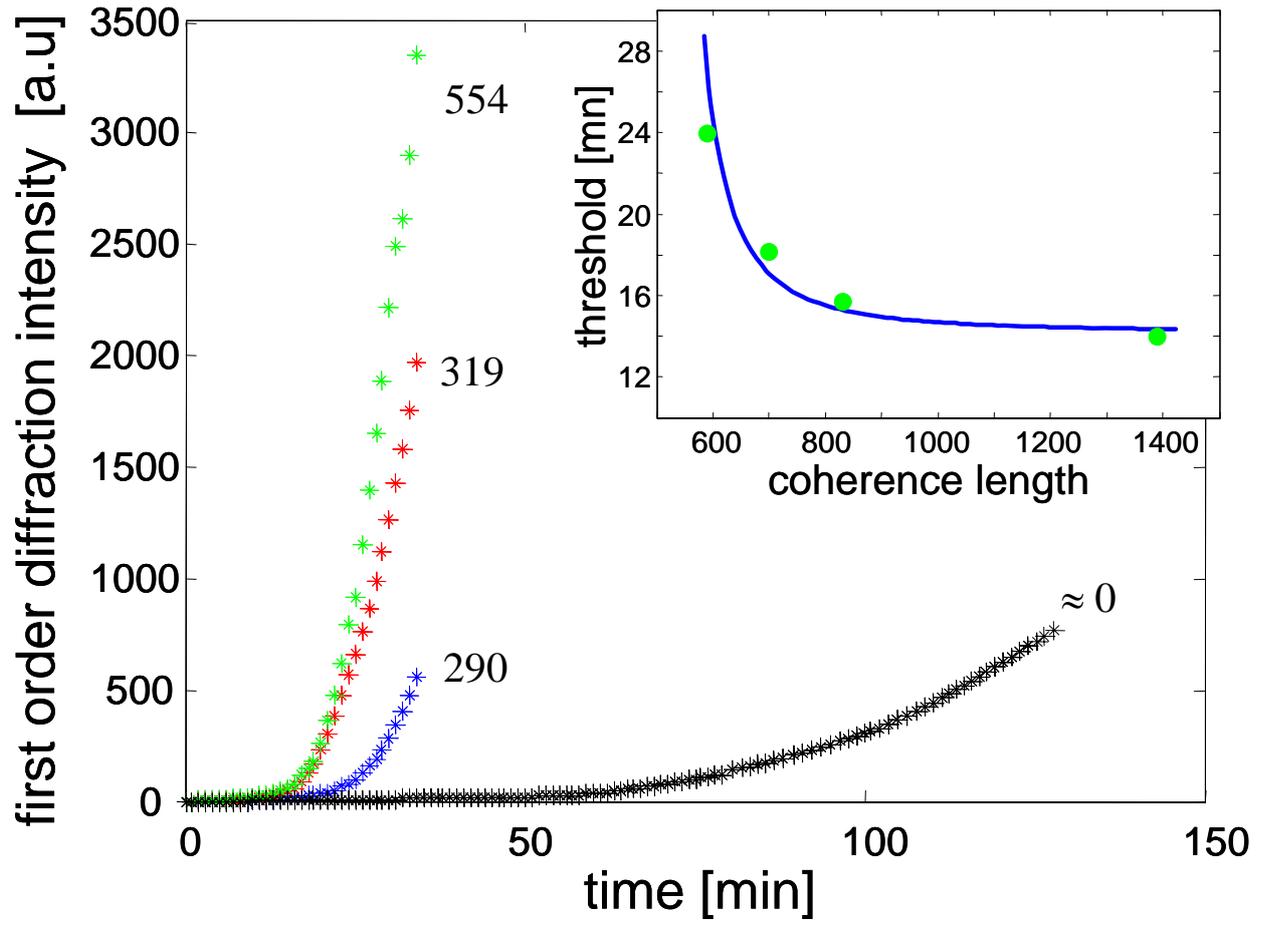

**Figure 2**

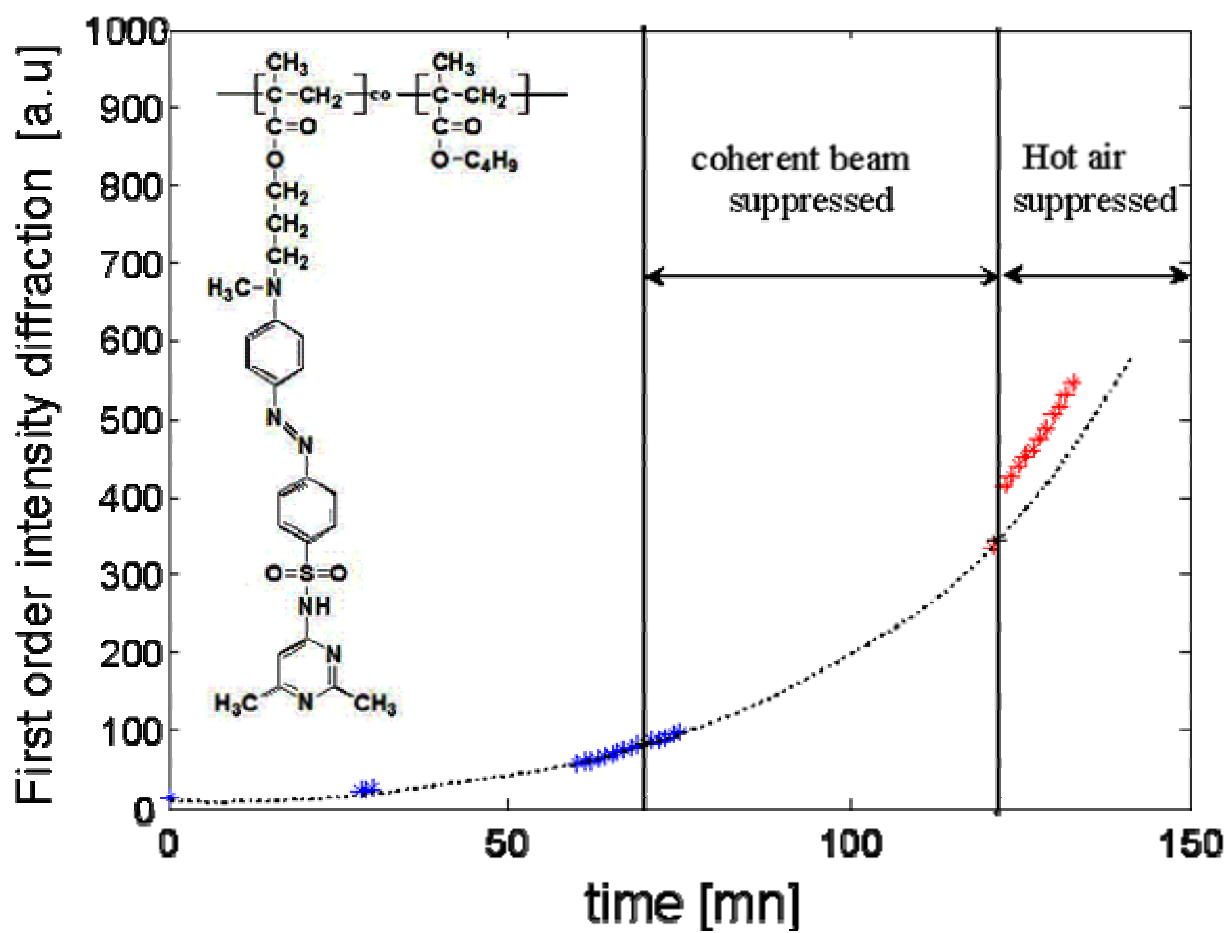

**Figure 3**

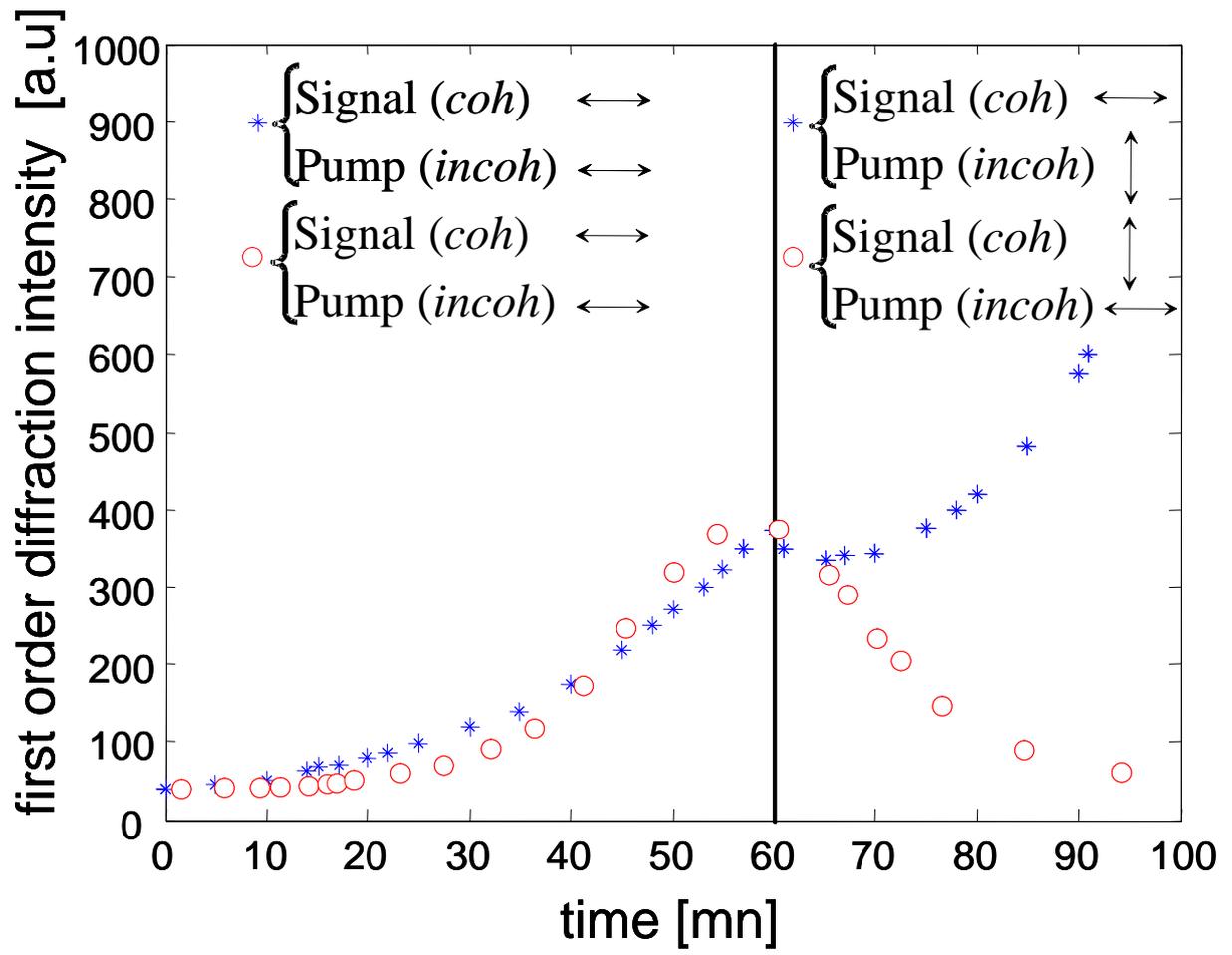

**Figure 4**